\documentclass[aps,prl, letter, twocolumn, superscriptaddress,showpacs,amsmath,amssymb]{revtex4}
\usepackage{graphicx}% Include figure files
\usepackage[normalem]{ulem}
\usepackage{color}
\definecolor{DarkBlue}{rgb}{0.1,0.1,0.5}
\definecolor{Red}{rgb}{0.9,0.0,0.1}
\definecolor{Green}{rgb}{0.0,0.99,0.0}

\begin{document}

\title{Real-time manifestation of strongly coupled spin and charge order parameters in stripe-ordered nickelates via time-resolved resonant x-ray diffraction}

\author{Y. D. Chuang}
\affiliation {Advanced Light Source, Lawrence Berkeley National Laboratory, Berkeley, CA 94720}

\author{W. S. Lee} \email{leews@stanford.edu}
\affiliation {Stanford Institute for Materials and Energy Sciences, SLAC National Accelerator Laboratory, Menlo Park, CA 94025, USA}

\author{Y. F. Kung}
\affiliation{Stanford Institute for Materials and Energy Sciences, SLAC National Accelerator Laboratory, Menlo Park, CA 94025, USA}

\author{A.P. Sorini}
\affiliation{Stanford Institute for Materials and Energy Sciences, SLAC National Accelerator Laboratory, Menlo Park, CA 94025, USA}
\affiliation{Physics Division, Lawrence Livermore National Laboratory, Livermore, CA 94550, USA}

\author{B. Moritz}
\affiliation{Stanford Institute for Materials and Energy Sciences, SLAC National Accelerator Laboratory, Menlo Park, CA 94025, USA}
\affiliation{Department of Physics and Astrophysics, University of North Dakota,Grand Forks, ND 58202, USA}
\affiliation{Department of Physics, Northern Illinois University, DeKalb, IL 60115, USA}

\author{R. G. Moore}
\affiliation{Stanford Institute for Materials and Energy Sciences, SLAC National Accelerator Laboratory, Menlo Park, CA 94025, USA}

\author{L. Patthey}
\affiliation{Stanford Institute for Materials and Energy Sciences, SLAC National Accelerator Laboratory, Menlo Park, CA 94025, USA}
\affiliation{Swiss Light Source, Paul Scherrer Institut, CH-5232 Villigen-PSI, Switzerland}

\author{M. Trigo}
\affiliation{Stanford Institute for Materials and Energy Sciences, SLAC National Accelerator Laboratory, Menlo Park, CA 94025, USA}
\affiliation{Stanford PULSE Institute, SLAC National Accelerator Laboratory, Menlo Park, CA 94025, USA}

\author{D. H. Lu}
\affiliation{Stanford Synchrotron Radiation Lightsource, SLAC National Accelerator Laboratory, Menlo Park, CA, 94025, USA}

\author{P. S. Kirchmann}
\affiliation{Stanford Institute for Materials and Energy Sciences, SLAC National Accelerator Laboratory, Menlo Park, CA 94025, USA}

\author{M. Yi}
\affiliation{Stanford Institute for Materials and Energy Sciences, SLAC National Accelerator Laboratory, Menlo Park, CA 94025, USA}

\author{O. Krupin}
\affiliation{European XFEL, Germany}
\affiliation{Linac Coherent Light Source, SLAC National Accelerator Laboratory, Menlo Park, CA 94720, USA}

\author{M. Langner}
\affiliation{Materials Science Division, Lawrence Berkeley National Laboratory, Berkeley, CA 94720, USA}

\author{Y. Zhu}
\affiliation{Materials Science Division, Lawrence Berkeley National Laboratory, Berkeley, CA 94720, USA}

\author{S. Y. Zhou}
\affiliation{Materials Science Division, Lawrence Berkeley National Laboratory, Berkeley, CA 94720, USA}

\author{D. A. Reis}
\affiliation{Stanford Institute for Materials and Energy Sciences, SLAC National Accelerator Laboratory, Menlo Park, CA 94025, USA}
\affiliation{Stanford PULSE Institute, SLAC National Accelerator Laboratory, Menlo Park, CA 94025, USA}

\author{N. Huse}
\affiliation{Max-Planck Research Group for Structural Dynamics, University of Hamburg, CFEL, Germany}

\author{J. S. Robinson}
\affiliation{Linac Coherent Light Source, SLAC National Accelerator Laboratory, Menlo Park, CA 94720, USA}

\author{R. A. Kaindl}
\affiliation{Materials Science Division, Lawrence Berkeley National Laboratory, Berkeley, CA 94720, USA}

\author{R. W. Schoenlein}
\affiliation{Materials Science Division, Lawrence Berkeley National Laboratory, Berkeley, CA 94720, USA}

\author{S. L. Johnson}
\affiliation{Swiss Light Source, Paul Scherrer Institut, CH-5232 Villigen-PSI, Switzerland}

\author{M. F\"{o}rst}
\affiliation{Max-Planck Research Group for Structural Dynamics, University of Hamburg, CFEL, Germany}

\author{D. Doering}
\affiliation{Engineering Division, Lawrence Berkeley National Laboratory, Berkeley, CA 94720, USA}

\author{P. Denes}
\affiliation{Engineering Division, Lawrence Berkeley National Laboratory, Berkeley, CA 94720, USA}

\author{W. F. Schlotter}
\affiliation{Linac Coherent Light Source, SLAC National Accelerator Laboratory, Menlo Park, CA 94720, USA}

\author{J. J. Turner}
\affiliation{Linac Coherent Light Source, SLAC National Accelerator Laboratory, Menlo Park, CA 94720, USA}

\author{T. Sasagawa}
\affiliation{Materials and Structures Laboratory, Tokyo Institute of Technology, Kanagawa 226-8503, Japan}

\author{Z. Hussain}
\affiliation {Advanced Light Source, Lawrence Berkeley National Laboratory, Berkeley, CA 94720}

\author{Z. X. Shen}\email{zxshen@stanford.edu}
\affiliation {Stanford Institute for Materials and Energy Sciences, SLAC National Accelerator Laboratory, Menlo Park, CA 94025, USA}

\author{T. P. Devereaux}\email{tpd@stanford.edu}
\affiliation {Stanford Institute for Materials and Energy Sciences, SLAC National Accelerator Laboratory, Menlo Park, CA 94025, USA}

\collaboration{Y. D. Chuang and W. S. Lee led the project and contributed equally to this work.}
\date{\today}% It is always \today, today,

\begin{abstract}
 We investigate the order parameter dynamics of the stripe-ordered nickelate,  La$_{1.75}$Sr$_{0.25}$NiO$_4$, using time-resolved resonant X-ray diffraction. In spite of distinct spin and charge energy scales, the two order parameters' amplitude dynamics are found to be linked together due to strong coupling. Additionally, the vector nature of the spin sector introduces a longer re-orientation time scale which is absent in the charge sector. These findings demonstrate that the correlation linking the symmetry-broken states does not unbind during the non-equilibrium process, and the time scales are not necessarily associated with the characteristic energy scales of individual degrees of freedom.
\end{abstract}

\pacs{Valid PACS appear here}% PACS, the Physics and Astronomy
\maketitle

In solids, the dynamics of electronic states are often represented in the energy domain.  For example, spin dynamics in ferromagnets are often characterized by the spin wave (i.e. magnon) whose bandwidth is proportional to the energy of spin exchange coupling. Charge dynamics are represented by characteristic energy scales in the single-particle excitations or the charge-charge correlation functions, such as energy gap, bandwidth, and collective mode energy. However, in complex materials, the strongly intertwined degrees of freedom can self-organize a large number of charges and spins into one or more collective broken symmetry states\cite{Anderson72,Laughlin00}, forming a rich phase diagram that is one of the hallmarks of strongly correlated materials\cite{Dagotto05}. In these cases, the dynamics of coexisting broken-symmetry states inevitably couple, and  the energy representation associated with a given degree of freedom may not provide a complete description of the dynamics of these collective states.

\begin{figure*} [t]
\includegraphics [clip, width=6.25 in]{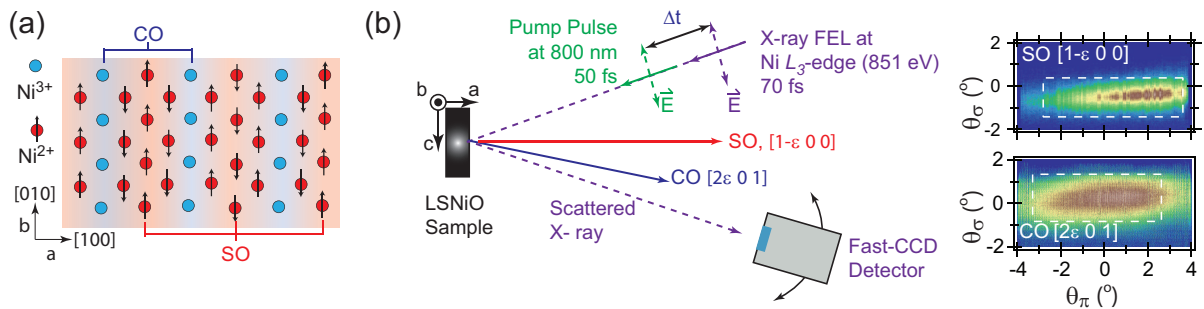}
\caption{\label{Fig1:ExperimentGeometry} (color online) (a) A schematic plot showing the charge (CO) and spin orders (SO) in a nickelate with a doping level of 0.25 holes/Ni. The black arrows represent the spin. (b) Top view of the experimental geometry. In the experiment, the sample was first pumped by an ultrafast (50 fs) 800 nm laser pulse, which we defined as occurring at time-zero; an XFEL pulse with a temporal duration shorter than 70 fs was introduced at a time delay $\Delta$t to map out the time evolution of the SO and CO resonant diffraction peaks. Images on the rightmost panels were produced by averaging over 30 - 40 single XFEL shot images to show the equilibrium state CO (bottom) and SO (top) resonant diffraction peak. The ordering vectors are listed in the images. $\epsilon$  is the incommensurability and is approximately 0.277(2$\pi$/a) for the measured sample. The horizontal axis $\theta_{\pi}$ is parallel to the scattering plane, cutting through the diffraction peaks in the [H 0 L] plane. The vertical axis $\theta_{\sigma}$, corresponding to the [0 K 0] direction, is perpendicular to the scattering plane. In this geometry, the diffraction peak shows an oval shape due to a much shorter correlation length along the c-axis than the [0 K 0] direction.}
\end{figure*}

Stripe-ordered nickelate\cite{Cava91} is a good example of materials, in which the dynamics of its broken-symmetry states is difficult to comprehend only using energy representation. In these striped nickelates, as shown in Fig. \ref{Fig1:ExperimentGeometry} (a), two broken-symmetry states of distinct degrees of freedom coexist \cite{Chen93,Schussler-Langeheine05,Yoshizawa00}: (1) charge order (CO), where doped charge carriers form one-dimensional charge density waves and break translational symmetry, and (2) spin order (SO), where antiferromagnetically ordered spin stripes are separated by the charge stripes and break both translational and rotational symmetry.  The known energy scales for the spin and charge degrees of freedom differ by at least an order of magnitude (20 meV for spin \cite{Woo05} and $>$200 meV \cite{Ido91,Katsufuji96} for charge); however, the significant energetic differences do not seem to be reflected in the emergent thermodynamic properties of CO and SO. More specifically, for a wide range of doping, the periodicity of the SO is always twice that of the CO, and their transition temperatures exhibit similar doping dependences\cite{Kajimoto03}, where the formation of the SO requires pre-existing CO. These observations are argued to provide supporting evidence that CO and SO are coupled strongly\cite{Zachar98}, which may be elusive in the energy scale description. Furthermore, such a coupling effect has not been explicitly revealed by experiments. Learning the dynamics of this stripe state could be relevant also to the physics of high T$_c$ superconducting cuprates, as stripes found in cuprates interact intimately with superconductivity\cite{Lake02,Kivelson03}.

Time-resolved pump-probe spectroscopy can offer a new perspective on these issues, because it can disentangle coupled degrees of freedom and the correlations between them by driving the system into a non-equilibrium state, and measuring the subsequent dynamics \cite{Giannetti11,Tomeljak09,Bigot09,Trigo10}. In our measurement, we directly monitor the CO and SO order parameters using time-resolved femtosecond resonant X-ray diffraction (RXD) at the Ni $L_3$-edge. RXD at the Ni $L_3$-edge significantly enhances the Bragg scattering from the modulation of Ni valence electrons at the CO and SO wave vectors, thus containing direct information about the respective order parameters\cite{Schussler-Langeheine05}. The intensity of resonant diffraction peak is proportional to both the square of the order parameters' amplitude and a Debye-Waller-like factor from the electronic phase fluctuation of the order parameters\cite{Overhauser71,Axe80,Chapman84}. Upon photo-excitation, we discover the CO is suppressed more than the SO, despite the fact that the CO is more robust thermodynamically. In the recovery process, both order parameters' amplitudes are locked to the same time scale, irrespective of the stark difference in their associated energy scales, demonstrating a real-time manifestation of the strong coupling between the CO and SO \cite{Zachar98}. We also find that the vector nature of SO results in a meta-stable state. Even though revealing the coupling and spin vector re-orientation dynamics is difficult in the energy domain, our findings show that these effects can be pronounced in the time domain. Results from a Gross-Pitaevskii time-dependent Ginzburg-Landau free energy model agree well with our experimental observations, lending further support to our conclusions.

Experiments were performed using the RSXS endstation \cite{Doering11} and X-ray free electron laser (FEL) at the SXR beamline\cite{Schlotter12} of the Linac Coherent Light Source (LCLS). Nickelate samples with a doping level of 0.25 hole/Ni were chosen for this measurement. The transition temperatures of CO and SO are approximately 110 K and 100 K, respectively. The samples were cut and polished to produce a mirror-like (1 0 0) surface. Orthorhombic notation is used to describe reciprocal space. The experimental geometry is illustrated in Fig. \ref{Fig1:ExperimentGeometry}(b) and the scattering plane lies in the a-c plane. Under this geometry, both the CO and SO can be measured on the same sample by rotating sample and detector angles respectively in the scattering plane. A X-ray pulse with a duration of 70 fs was aligned co-linearly with the optical pump laser (wavelength 800 nm, 50 fs). The overall experimental temporal resolution was approximately 0.4 ps, which is limited by the accuracy of synchronizing the pump laser and X-ray FEL probe pulse. More information about the experimental method can be found in the Supplementary Materials.

Figure \ref{Fig2:SOCOSupression}(a) shows the time traces of integrated peak intensity for both CO and SO. Upon photo-excitation, the intensities of both CO and SO drop in proportion to the pump excitation density, and after reaching a minimum, begin to recover toward their original values. We first note that the peak position and width remain unchanged for the CO\cite{Lee12} and SO (Supplementary Materials) throughout the measurement time window. As discussed previously\cite{Lee12}, the invariance in the peak position and width suggests that no topological defects are created to change the periodicity and/or the correlation lengths in the photo-induced transient states. This finding is in stark contrast to the thermal evolution, where both periodicities and coherence lengths exhibit continuous changes with the temperature\cite{Chen93}. Therefore, the photo-induced transient state cannot be explained simply by thermal processes at elevated electronic and lattice temperatures, which is further supported by the lattice dynamics shown by (0,0,2) Bragg peak in Fig. \ref{Fig3:TImeScale}(b), as will be discussed later.

Interestingly, following photo-excitation, the integrated intensity of the CO peak exhibits a higher degree of initial suppression than that of the SO peak. This is further highlighted in Fig. \ref{Fig2:SOCOSupression}(b) by plotting the maximal change of normalized intensity near time zero (i.e. the magnitude of the drop in the time trace) as a function of photo-excitation density. Irrespective of the excitation density, $|\Delta I_{max}/I_0|$ for the CO is always larger than that for the SO. Notably, at the largest photo-excitation densities, the CO is completely suppressed below the detection threshold ($\Delta I_{max}/I_0$  = -1.0, shaded area in Fig. \ref{Fig2:SOCOSupression}b) yet the SO remains detectable. This intriguingly contrasts the behavior in thermal equilibrium, where the CO is known to be more robust against the thermal fluctuations because of higher transition temperature. Thus, the photo-induced electronic transitions (inter- and intra-Ni ions \cite{Ido91}) can create an unbalanced CO and SO transient state because of stronger photon-charge coupling than photon-spin coupling (see Supplementary Materials for a simple cluster model simulation). This unbalanced transient state is inaccessible from thermal equilibrium, which provides a unique stage for investigating the emergent dynamics of CO and SO during the recovery process.

\begin{figure}
\includegraphics [clip, width=3.25 in]{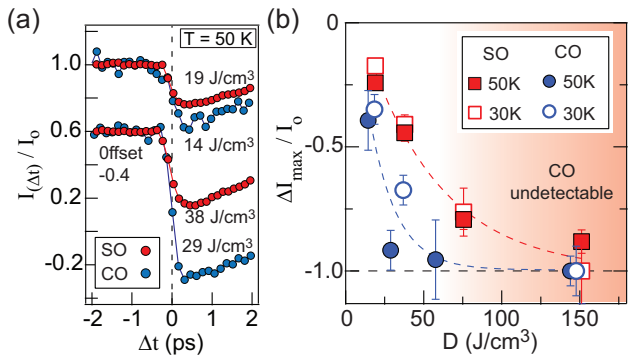}
\caption{\label{Fig2:SOCOSupression} (color online) (a) Time traces showing the normalized CO (blue) and SO (red) peak intensities at selected photo-excitation densities. $\Delta t$ is the time delay between the pump laser pulse and X-ray probe pulse. The peak intensities are calculated by integrating the false-color images over the white box in Fig. \ref{Fig1:ExperimentGeometry}(c). The curves measured at higher photo-excitation densities are offset by -0.4 for clarity. (b) A summary plot of the maximal change in the normalized intensity as a function of photo-excitation density. The error bars are estimated from the instrumental noise in each time trace and the dashed lines are guides-to-the-eye. The shaded area indicates the region where the transient CO signal is below the detection limit.}
\end{figure}

Figure \ref{Fig3:TImeScale}(a) shows the time traces of the CO and SO recorded at similar photo-excitation densities. The CO and SO recover at similar rates within the first few picoseconds. Beyond that, CO continues to recover at a slower rate while SO settles into a long-lived meta-stable state. We further quantify the recovery time scales by fitting the time traces with a two-time-scale model. This model, previously used to extract the time scales for CO\cite{Lee12}, also gives a reasonable fit to the SO time traces (Fig. \ref{Fig3:TImeScale}(b)). The extracted time scales for both CO and SO are summarized in Fig. \ref{Fig3:TImeScale}(c). Remarkably, the time scales of the faster dynamics are comparable and lie within the range of 2 - 4 picoseconds. In contrast, even with much larger error bars, the time scales for slower dynamics clearly show an order of magnitude difference between CO and SO, confirming the apparent meta-stability of SO. One might speculate whether the observed SO and CO dynamics are dragged by the lattice dynamics, which could be described using an effective lattice temperature scenario. To resolve this speculation, we demonstrate the dynamics of the (0,0,2) lattice Bragg peak also in Fig. \ref{Fig3:TImeScale}(b). It is found that the lattice dynamics is slow without any recovery during the time window of our data ( $\Delta t < $  25 ps). This is in stark contrast with the rich time scales exhibited in the SO and CO dynamics. This observation, together with the aforementioned lack of change of order parameters' coherent length and periodicity, asserts that the observed SO and CO dynamics is indeed primarily electronic, not determined or bottlenecked by the lattice dynamics.

\begin{figure*}
\includegraphics [clip, width=6.25 in]{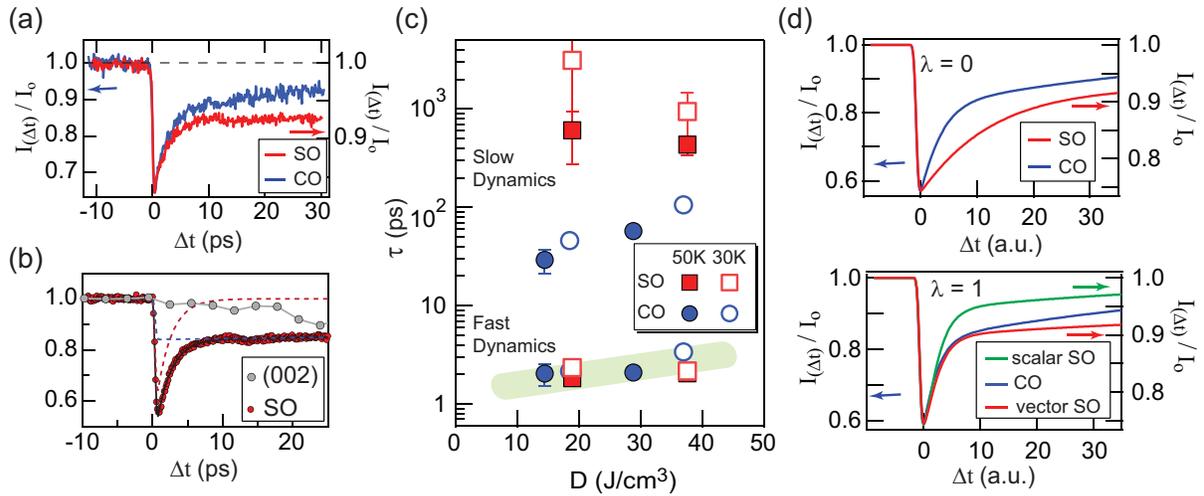}
\caption{\label{Fig3:TImeScale} (color online) (a) Time traces of the normalized CO (blue curve, left axis) and SO (red curve, right axis) peak intensities at a photo-excitation density of approximately 19 J/cm$^3$ at 30 K. (b) Two-time-scale model fitting (black curve) to the SO time trace (red curve). The fitted initial fast (red dashed curve) and slow (blue dashed curve) dynamics are also plotted. The data of (0,0,2) Bragg peak are also included to illustrate a very different lattice dynamics than those exhibited in CO and SO. The demonstrated SO and (0,0,2) Bragg peak data were taken at 50 K with an excitation densities of 38 J/cm$^3$.(c) A summary plot of the fast and slow time scales of the CO\cite{Lee12} (blue circles) and SO (red squares) as a function of photo-excitation density. The green shaded area is a guide-to-the-eye to highlight a comparable fast time scale for the CO and SO. The dynamics at 30 K (open markers) and 50 K (filled markers) are essentially the same within the experimental uncertainties. The error bars are determined by 95\% confidence interval of fitting. The long-lived meta-stable state of SO dynamics has a life time much longer than the recorded time window, yielding a large error bar. (d) Simulated time traces of the normalized CO and SO peak intensities from time-dependent Ginzburg-Landau theory. The upper (lower) panel shows the results when the SO-CO coupling is zero (of the order 1). In the lower panel, the trace labeled with ``vector SO" includes the SO vector re-orientation effect.}
\end{figure*}

\begin{figure}
\includegraphics [clip, width=3.25 in]{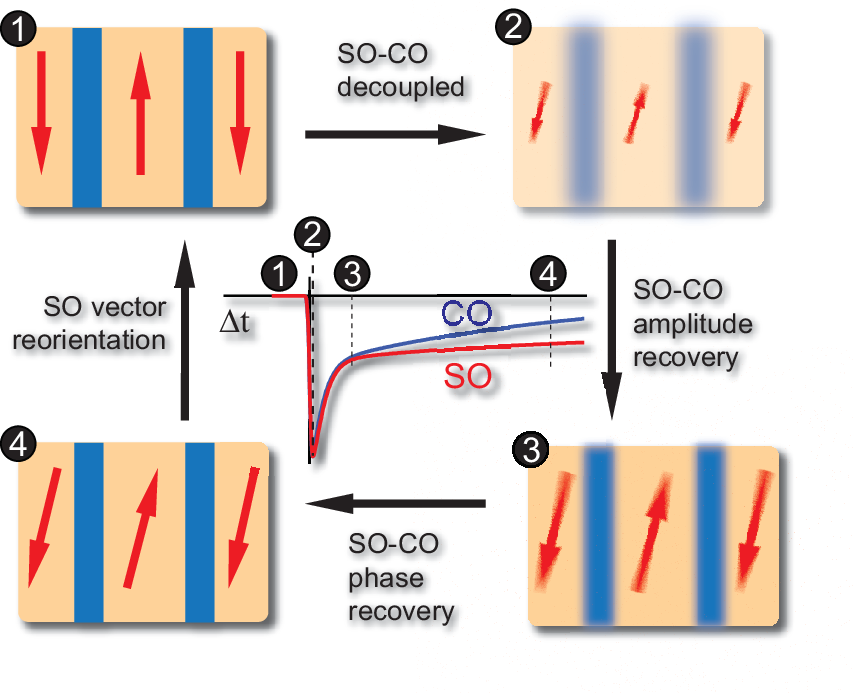}
\caption{\label{Fig4:Summary} (color online) Real-space cartoon snapshots of the behavior of stripes at four different stages (1-4) of the pump-probe experiment. The yellow (blue) stripes represent the SO (CO), with red arrows indicating the spin orientations. Amplitude of the order parameters is represented by the color depth, as well as the length of the arrows for SO. The blurring effect is used to signal the presence of phase excitations. }
\end{figure}

These distinct time scales speak for the dynamics of the order parameters. While the relaxation process during the recovery can be complex at microscopic levels, RXD signal is only sensitive to the long range order; therefore, the times scales exhibited in the time-resolved RXD measurement speak for order parameters' dynamics, regardless of the underlying microscopic relaxation mechanisms. In our recent work on the same compound, we have established that the fast dynamics of photo-excited CO are related to the recovery of the order parameter's amplitude, whereas the slower dynamics are determined by the recovery of the order parameter's phase, i.e. the relaxation of incoherent phasons induced by photo-excitations\cite{Lee12}. For the SO, similarly, there are three candidates for the fast dynamics: recovery of amplitude, phase, and the SO orientation. Since the recovery of SO's phase and orientation depends on the weak SO-phason-lattice coupling and spin-orientation-lattice coupling, respectively, these two dynamics are expected to be slower than the amplitude dynamics. Thus, the observation of comparable time scales for the fast dynamics of the CO and SO indicates that recovery of the two order parameters' amplitudes locks together, despite the distinct suppression of the two orders.  This observation is highly non-intuitive, considering that the magnitudes of the underlying energy scales are fundamentally different. Namely, the SO should be closely related to the spin gap and spin exchange energy\cite{Woo05} ($\sim$ 20 meV), while a balance between the Coulomb and electronic kinetic energy ($>$200 meV\cite{Ido91,Katsufuji96}) should drive the CO. Thus, distinct time scales for the the two orders are expected, opposite to our experimental observations. Apparently, our time scale analysis reveals a hidden dynamics that is undiscernable by the known energy scales associated with spin and charge degrees of freedom.

We argue that this hidden effect is the strong dynamical coupling between the CO and SO, which is further supported by the calculation using a Gross-Pitaevskii time-dependent Ginzburg-Landau free energy model (Supplementary Material). In this model, the free energy parameters are set such that the CO and SO transitions remain second order. Phenomenological damping coefficients are used to set the rates of energy dissipation for order parameters' amplitudes and phases. Since phase excitation is weakly coupled to the lattice, a much slower energy dissipation rate is assigned to the  phase excitations. (However, we note that varying the damping factors themselves does not change the qualitative behavior, as shown in the Supplementary Material.) The initial suppression of CO and SO at time zero are adjusted close to the experimental data for direct comparison. In the absence of CO-SO coupling (Fig. \ref{Fig3:TImeScale}(d), upper panel), the CO and SO time traces exhibit distinct dynamics in accordance with their intrinsic relevant energy scales. Switching on strong CO-SO coupling (on the order of 1) locks the real-space orderings, initial recovery time scales, i.e. the amplitude dynamics (blue and green curves in lower panel of Fig. \ref{Fig3:TImeScale}(d)) in agreement with our experimental observations (Fig. \ref{Fig3:TImeScale}(a) and \ref{Fig3:TImeScale}(c)). Furthermore, the CO and SO phases slowly recover also in a coupled manner  as shown in Fig. \ref{Fig3:TImeScale}(d). In short, the CO-SO coupling reduces four independent time scales to two fundamental time scales, one each for phase and amplitude dynamics.

However, this CO-SO coupling alone fails to reproduce the observed meta-stability in the SO peak intensity at long time. Undoubtedly, the missing ingredient is a unique property of SO since such meta-stability is absent in the CO dynamics. We also note that the SO diffraction peak intensity should be modulated by a light scattering geometric factor, $|(\epsilon ' \times \epsilon)\cdot m|^2$, where $m$, $\epsilon$ and $\epsilon '$ represent the unit vector of SO, and polarizations of the scattered and incident X-rays, respectively\cite{Hill96}. As a consequence, any misalignment of the SO vector from its original direction would reduce the SO peak intensity in addition to the reduction from SO's amplitude and phase fluctuations. Furthermore, the equilibrium orientation of SO is known to change with temperature \cite{Freeman}, implying its glassy behavior and sensitivity to perturbations. Thus, when driven out of equilibrium, the time the SO needs to re-align can be long, obscuring SO's phase recovery time scale extracted in our experiments. Finally we note that by including this long-lived meta-stable state, the calculated time traces for SO and CO diffraction peak intensity (red and blue curves in the lower panel of Fig. \ref{Fig3:TImeScale}(d)) show satisfactory agreement with experimental observation (Fig. \ref{Fig3:TImeScale}(a)).

The cooperative dynamics are summarized in Fig. \ref{Fig4:Summary}. Initially, the photo-excitation not only reduces the CO and SO amplitudes, but also perturbs their phases and SO's vector orientation. Although the pump pulse excitation immediately unbinds the CO-SO coupling, the evolution of CO and SO collective dynamics - the phase and amplitude - remain locked during all time of the recovery, reflecting the scalar nature of both order parameters. Only the spin vector evolves toward equilibrium at its own time scale. The persistence of the correlations linking broken symmetry states in non-equilibrium process could be generic to other broken symmetry states in other correlated materials when driven-out-of-equilibrium.

This research was supported by the U.S. Department of Energy, Office of Basic Energy Sciences, Division of Materials Sciences and Engineering under Contract No. DE-AC02-76SF00515, SLAC National Accelerator Laboratory (SLAC), Stanford Institute for Materials and Energy Science (W.S.L., R.M., L.P., M.T, D.A.R, Y.F.K., A.P.S., B.M, T.P.D., Z.X.S), SLAC Stanford Synchrotron Radiation Lightsource (D.H.L.), SLAC Stanford PULSE Institute (M.T., D.A.R.) and under contract number DE-AC02-05CH11231 Lawrence Berkeley National Laboratory (LBNL) Advanced Light Source (Y.D.C., Z.H.), LBNL Materials Sciences Division (Y.Z., S.Z., R.A.K., R.W.S.), and LBNL Engineering Division (D.D., P.D.). P.S.K acknowledges support by the Alexander-von-Humboldt Foundation through a Feodor-Lynen scholarship. Y.F.K. was supported by the Department of Defense (DoD) through the National Defense Science and Engineering Graduate Fellowship (NDSEG) Program. The SXR Instrument at LCLS is funded by a consortium whose membership includes LCLS, Stanford University - SIMES, LBNL, University of Hamburg through the BMBF priority program FSP 301, and the Center for Free Electron Laser Science (CFEL).


\begin{thebibliography}{99}

\bibitem{Anderson72}
P. Anderson, Science \textbf{177}, 393 (1972).

\bibitem{Laughlin00}
R. B. Laughlin, PNAS \textbf{97}, 28 (2000).

\bibitem{Dagotto05}
E. Dagotto, Science \textbf{309}, 257 (2005).

\bibitem{Cava91}
R. J. Cava \emph{et al.}, Phys. Rev. B \textbf{43}, 1229 (1991).

\bibitem{Chen93}
C. H. Chen, S.-W. Cheong, and A. S. Cooper, Phys. Rev. Lett \textbf{71}, 2461 (1993).

\bibitem{Schussler-Langeheine05}
C. Sch$\ddot{u}$ssler-Langeheine \emph{et al.}, Phys. Rev. Lett \textbf{95}, 156402 (2005)

\bibitem{Yoshizawa00}
H. Yoshizawa \emph{et al.}, Phys. Rev. B \textbf{61}, R854 (2000).

\bibitem{Woo05}
Hyungje Woo \emph{et. al.} Phys. Rev. B \textbf{72}, 064437 (2005).

\bibitem{Ido91}
T. Ido, K. Magoshi, H. Eisaki, and S. Uchida, Phys. Rev. B \textbf{44}, 12094 (1991).

\bibitem{Katsufuji96}
T. Katsujuji \emph{et al.}, Phys. Rev. B \textbf{54}, 14230 (1996).

\bibitem{Kajimoto03}
R. Kajimoto, K. Ishizaka, H. Yoshizawa, and Y. Tokura, Phys. Rev. B \textbf{67}, 014511 (2003).

\bibitem{Zachar98}
O. Zachar, S. A. Kivelson, and V. J. Emery, Phys. Rev. B \textbf{57}, 1422 (1998).

\bibitem{Lake02}
B. Lake \emph{et al.}, Nature \textbf{455}, 299 (2002).

\bibitem{Kivelson03}
S. A. Kivelson, I. P. Bindloss, E. Fradkin, V. Oganesyan, J. M. Tranquada, A. Kapitulnik, and C. Howald, Rev. Mod. Phys. \textbf{75}, 1201 (2003).

\bibitem{Giannetti11}
C. Giannetti \emph{et al.}, Nature Commun. \textbf{2}, 353 (2011).

\bibitem{Tomeljak09}
A. Tomeljak, H. Schafer, D. Stadter, M. Beyer, K. Biljakovic, and J. Demsar, Phys. Rev. Lett. \textbf{102}, 066404 (2009).

\bibitem{Bigot09}
Jean-Yves Bigot, Mircea Vomir, and Eric Beaurepaire, Nature Phys. \textbf{5}, 515 (2009).

\bibitem{Trigo10}
M. Trigo and D. Reis, MRS Bulletin \textbf{35}, 514 (2010).

\bibitem{Overhauser71}
A. W. Overhauser, Phys. Rev. B \textbf{3}, 3173 (1971).

\bibitem{Axe80}
J. D. Axe, Phys. Rev. B \textbf{21}, 4181 (1980).

\bibitem{Chapman84}
L. D. Chapman and R. Colella, Phys. Rev. Lett. \textbf{52}, 652 (1984).

\bibitem{Ichikawa11}
H. Ichikawa \emph{et al.}, Nature Mater. \textbf{10}, 101 (2011).

\bibitem{Matteo07}
M. Rini \emph{et al.}, Nature \textbf{449}, 72 (2007).

\bibitem{Fausti11}
D. Fausti \emph{et al.}, Science \textbf{331}, 189 (2011).

\bibitem{Doering11}
D. Doering \emph{et al.}, Rec. Sci. Instrum. \textbf{82}, 073303 (2011).

\bibitem{Schlotter12}
W. F. Schlotter \emph{et al.} Rev. Sci. Instrum. \textbf{83}, 043107 (2012).


\bibitem{Lee12}
W. S. Lee \emph{et al.}, Nature Commun. \textbf{3}, 838 (2012).

\bibitem{Hill96}
J. P. Hill and D. F. McMorrow, Acta. Cryst. A \textbf{52}, 236 (1996).

\bibitem{Freeman}
P. G. Freeman, A. T. Boothroyd, D. Prabhakaran, M. Enderle, C. Niedermayer, Phys. Rev. B \textbf{70}, 024413 (2004).

\end{thebibliography}
\end{document}